# Mixed Neural Network Approach for Temporal Sleep Stage Classification

Hao Dong, Akara Supratak, Wei Pan, Chao Wu, Paul M. Matthews and Yike Guo*

*Abstract*—This paper proposes a practical approach to addressing limitations posed by using of single-channel electroencephalography (EEG) for sleep stage classification. EEG-based characterizations of sleep stage progression contribute the diagnosis and monitoring of the many pathologies of sleep. Several prior reports explored ways of automating the analysis of sleep EEG and of reducing the complexity of the data needed for reliable discrimination of sleep stages at lower cost in the home. However, these reports have involved recordings from electrodes placed on the cranial vertex or occiput, which are both uncomfortable and difficult to position. Previous studies of sleep stage scoring that used only frontal electrodes with a hierarchical decision tree motivated this paper, in which we have taken advantage of rectifier neural network for detecting hierarchical features and long short-term memory (LSTM) network for sequential data learning to optimize classification performance with single-channel recordings. After exploring alternative electrode placements, we found a comfortable configuration of a single-channel EEG on the forehead and have shown that it can be integrated with additional electrodes for simultaneous recording of the electrooculogram (EOG). Evaluation of data from 62 people (with 494 hours sleep) demonstrated better performance of our analytical algorithm than is available from existing approaches with vertex or occipital electrode placements. Use of this recording configuration with neural network deconvolution promises to make clinically indicated home sleep studies practical.

## I. INTRODUCTION

People spend approximately one-third of their life sleeping and sleep plays an important role in physiological homeostasis. Sleep related disorders, such as sleep apnea, insomnia, narcolepsy, reduce the quality of life for the large numbers of people who are affected. As much as 33% of general population reports suffering from insomnia [1]. For accurate diagnosis of sleep disorders, all-night polysomnographic (PSG) recording including an electroencephalogram (EEG), electrooculogram (EOG) and electromyogram (EMG), followed by expert manual scoring of sleep stages and their progression according to standard guidelines is needed [2], [3]. High costs and limited availability of specialized facilities limit their use.

As home sleep monitoring and automatic sleep stage scoring could reduce costs and increase access to diagnostic sleep studies, there has been interest in coupling the development of simple, wearable EEG recording devices with automated sleep stage classification. Three main challenges to the automatic sleep stage classification have been identified:

*Challenge 1. Heterogeneity.* People have different cranial structures and vary demographically and physiologically in ways that influence EEG patterns in sleep. For example, about 10% people do not generate alpha rhythm during stage W, and a further 10% generate only a limited alpha rhythm [3]. For these subjects, American Association of Sleep Medicine (AASM) guidelines suggest use of alternative criteria for classification of stages W and N1.

*Challenge 2. Temporal Pattern Recognition.* Scoring sleep stage is a sequential problem [3], as sleep stage scoring depends not only on temporally local features, but also on prior epochs time. For example, the onset of stage N2 depends on whether K complex or sleep spindles occurs early or in the last half of the previous epoch [3]; stage N2 can be classified even without K complexes or sleep spindles. Rapid eye movement sleep (REM) classification also depends on the features from prior EEG epochs, e.g., an epoch can be scored as REM, even in the absence of rapid eye movements, if the chin EMG tone is low and at low amplitude and there is mixed frequency EEG activity without K complexes or sleep spindles.

*Challenge 3. Comfort.* Previous reports [4], [5], [6], [7] described home sleep EEG recording with montages including central, occipital and parietal electrodes [1], which better detect sleep spindles, vertex shape waves and alpha rhythm than do frontal electrode (Table II). However, these EEG positions demand placement of the electrodes in hairy regions of scalp, demanding careful placement and adhesive paste to minimize movement related noise, and can lead to limitation of head movement and discomfort during sleep [8].

A recent study [4] evaluated accuracy of classification with Fpz-Cz channel [10], using a Complex Morlet wavelets transform for feature extraction and Stacked Sparse Autoencoders for classification. This showed that including the features from neighboring epochs can improve the classification performance. It also highlighted a bias towards misclassification of epochs as the overall most frequently occurring class (stage N2) because of the inherent imbalance in occurrence of the different sleep stages. To solve the imbalance problem, the author used a down-sampling method to generate new, balanced dataset in which every sleep stage is equally represented. To use more information from the original training dataset,

H. Dong, A. Supratak, W. Pan, C. Wu and Y. Guo are with the Department of Computing, Imperial College London, London, SW7 2AZ, UK (e-mail: hao.dong11, as12212, w.pan11, chao.wu, y.guo@ic.ac.uk)

PM, Matthews is with the Division of Brain Sciences in the Department of Medicine, Imperial College London, London, SW7 2AZ, UK (e-mail: p.matthews@ic.ac.uk)

* Corresponding author

[1]The location of scalp electrodes for sleep scoring is shown in Fig. 1, following the international 10/20 system, in which each site has a letter to identify the lobe and a number to identify the hemisphere location.

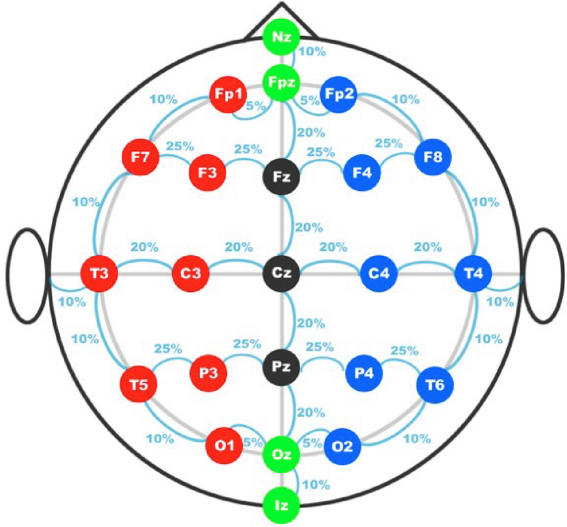

Fig. 1. The international 10/20 EEG system [9]. The letters F, T, C, P and O refer to frontal, temporal, central, parietal and occipital lobes placements, respectively. The even numbers refer to electrode positions on the right hemisphere, the odd numbers refer to electrode positions on the left hemisphere and the 'z' refers to electrode placement on the mid line of the head. Additionally, A1 and A2 define position on the left and right earlobes, respectively.

the authors generated new "balanced" datasets and trained individual networks using data from each of their subjects (ensemble learning). However, to obtain one prediction, feed-forward propagations on each of the individual networks are required. This is inefficient, although it improves accuracy. Here, we have chosen an alternative "ensemble learning" method that gains in efficiency by using dropout [11].

Other reports describe ways in which classification accuracy can be improved by supplementing data with EEG recordings from central, occipital or parietal electrode. For example, [5] evaluated a method using a C3-A1 channel. In [6], alternative approaches using a C3-A2 and the Pz-Oz channels were described. Here, the author used multi-scale entropy (MSE) and autoregressive (AR) models as features, and then trained a linear discriminant analysis (LDA) model as a classifier. In [7], classification was based on the Cz-Pz channel. However, all these three studies evaluated their methods without any type of cross-validation and one [5] trained the classifier using signals from all pf the subjects, which meant that training and testing data were not independent.

To address problem of wearer comfort, time-frequency domain features extracted from Fp1-Fp2 EEG and left-right EOG channels have been investigated in [12], [13], [14], [15], [16], [17]. These features were fed into a hierarchical decision tree [12], [13], [14], [15], [17], or a structured hierarchical SVM [16] classifier to determine different their corresponding sleep stages. However, these approaches still require prior knowledge to define the structure of the decision trees. This makes the primary distinction with our method, as we aim to utilize a long short-term memory (LSTM) to automatically learn the scoring strategy instead of manually defining rules in the decision tree. TABLE XI-XVI show the results of five representative studies.

In our approach, we propose use of a Mixed Neural Network (MNN) to solve both the population heterogeneity and temporal pattern recognition problems. Our MNN is composed of a rectifier neural network which suitable for detecting naturally sparse patterns [18], and a long short-term memory (LSTM) for detection of temporally sequential patterns [19]. We will describe the details in Section II-C. For signal recording, we propose a novel configuration that combines a low frontal electrode for EEG signal detection with another electrode for electrooculography (EOG). During periods without eye movement, the latter electrodes act as reference electrodes (analogous to A1 and A2). Through the full course of the study, the EOG provides additional information for sleep scoring by detecting eye movements [20], [21].

TABLE I
PERFORMANCE OF LITERATURE. OVERALL ACCURACY (ACC), MACRO
F1-SCORE (MF1), F1-SCORE (F1). THE LETTERS F, T, C, P AND O REFER
TO FRONTAL, TEMPORAL, CENTRAL, PARIETAL AND OCCIPITAL LOBES
PLACEMENTS, RESPECTIVELY.

| Method | F1 | | | | | MF1 | ACC |
|---|---|---|---|---|---|---|---|
| | W | N1 | N2 | N3 | REM | | |
| Fpz-Cz [4] | 71.58 | 47.04 | 84.60 | 84.03 | 81.40 | 73.73 | 78.94 |
| Cz-Pz [5] | 91.45 | 47.62 | 82.59 | 74.21 | 77.81 | 74.74 | 82.57 |
| C3-A2/Pz-Oz [6] | 93.62 | 15.29 | 78.25 | 71.45 | 81.96 | 68.11 | 83.60 |
| Cz-Pz [7] | 85.95 | 20.86 | 84.78 | 84.28 | 85.95 | 72.36 | 82.93 |
| Fp1-Fp2 [14] | 74.82 | 47.04 | 86.94 | 87.89 | 86.31 | 76.59 | 81.65 |

## II. METHODOLOGY

### A. Sleep stage standards

There are two standards commonly used to define sleep stages: the Rechtschaffen and Kales (R&K) [2], and that developed by the American Academy of Sleep Medicine (AASM) [3]. The AASM standard adopted for this paper, classifies sleep into 5 different stages with one awake stage (W), three sleep stages (N1, N2, N3) corresponding to different depths of sleep, and one rapid eye movement stage (REM). Table II summarizes the waves and events of EEG during sleep included in the AASM standards. Each sleep staging decision is based on a 30 (or 20) seconds window of the physiological signals called an EEG epoch.

### B. Features selection based on sleep physiology

The physiological features of sleep EEG can be typically characterized either in the time or frequency-domain (Table II).

*1) Spectral power of frequency bands:* Our approach uses time-frequency analysis to extract feature from each EEG epoch (Table III). We have chosen to use a conventional Fourier transform over other methods (e.g., complex Morlet wavelets) to allow us to strictly follow the AASM standards and then take advantage of neural network to further extract feature.

A short-time Fourier transform (STFT) [23] is used to extract temporal features. STFT extracts the frequency and phase content of a signal as it changes over time to generate a spectrogram. STFT has three parameters: the sliding window size, the overlap percentage, and the window function. The sliding window size defines the time interval of an EEG segment and controls the trade-off between frequency and temporal precision. e.g., increasing the window size will increase frequency precision, but decrease temporal precision. We allow segments to overlap to reduce artifacts at the boundaries of adjacent windows. Increasing the degree of overlap will decrease artifacts, but will also lead to higher computation costs. Window function is used to reduce the spectral leakage at the boundary of a sliding window.

TABLE II
EEG WAVES AND EVENTS DURING SLEEP

| Event | Frequency/Duration | Best Location | Related Stage |
|---|---|---|---|
| Alpha rhythm [3] | 8-13 Hz | Occipital lobe | W |
| Eye blinks | 0.5-2 Hz | EOG channel | W |
| Reading eye movement | uncertain | EOG channel | W |
| Rapid eye movement [20] | 0.5-2 Hz < 0.5 s | EOG channel | W, REM |
| Slow eye movement [21] | 0.1-1 Hz > 0.5 s | EOG channel | N1 |
| Low amplitude, mixed frequency activity [22] | 4-7 Hz | Frontal and Central lobe | N1 |
| Vertex shape waves | < 0.5 s | Central lobe | N1 |
| K complex | 1.6-4 Hz >= 0.5 s | Frontal lobe | N2 |
| Sleep spindle | 11-16 Hz >= 0.5 s | Central lobe | N2, N3 |
| Major body movement | > 15 s | All channels | All stages |
| Slow wave activity | 0.5-2 Hz > 75 $uV$ | Frontal lobe | N3 |
| Low chin EMG tone | 15-30 Hz | —— | REM |
| Sawtooth waves | 2-6 Hz | Central lobe | REM |
| Transient muscle activity | < 0.25 s | —— | REM |
| Arousal | 15-30 Hz | —— | N1 |

TABLE III
EEG FEATURES EXTRACTION

| Feature | Purpose | Related Stage |
|---|---|---|
| Maximum and minimum signal amplitude over the entire epoch | Capture the major body movement and peaks | All stages |
| Shannon entropy over the entire epoch | Capture the amplitude of vibration | All stages |
| Maximum and minimum signal amplitude using a sliding window | Capture the major body movement and peaks | All stages |
| 0.06-0.1 Hz using a sliding window | Capture the slow eye movement | N1 |
| 0.1-0.3 Hz using a sliding window | Capture the slow eye movement | N1 |
| 0.3-0.5 Hz using a sliding window | Capture the slow eye movement | N1 |
| 0.5-1 Hz using a sliding window | Capture the slow eye movement | N1 |
| 0.5-2 Hz using a sliding window | Capture the eye blink, rapid eye movement and slow wave activity | W, REM, N3 |
| 1.6-4 Hz using a sliding window | Capture the K complex | N2 |
| 3-4.5 Hz using a sliding window | Capture the hypersynchrony for children | N1 |
| 4-7 Hz using a sliding window | Capture the low amplitude, mixed frequency activity and rhythmic anterior theta activity (5-7 Hz, for children) | N1 |
| 8-13 Hz using a sliding window | Capture the alpha rhythm | W |
| 11-16 Hz using a sliding window | Capture the sleep spindle | N2, N3 |
| 15-30 Hz using a sliding window | Capture the low chin EMG tone | REM |
| For the frequency-band power capture using a sliding window, calculate it's maximum, minimum, mean, median and standard derivation | Capture the occasionally and continuous features | All stages |
| 0.06-0.1 Hz over the entire epoch | Capture the slow eye movement | N1 |
| 0.1-0.3 Hz over the entire epoch | Capture the slow eye movement | N1 |
| 0.3-0.5 Hz over the entire epoch | Capture the slow eye movement | N1 |
| 0.5-1 Hz over the entire epoch | Capture the slow eye movement | N1 |

In our experiment, the STFT was used to divide the 30 seconds EEG epochs into shorter segments of 5 seconds with serial overlaps of 70%. Each of the short EEG segments was windowed by hamming window function. The window length, window overlap, and window function are selected based on the performance in our experiment. For the window length, we tried from 1 to 8 seconds, and we found that the window length of 5 gave us the best performance. For the window overlap, we tried with 50%, 60%, 70%, 80% and 90%. We found that 70% gave us the best performance. We explored the use of both Hamming and Hanning windows and did not observe any performance distinction between the two.

After generating a spectrogram by using STFT, the spectral power of different sub-bands were calculated by summing up the amplitude values in each segment to define the power spectrum density (PSD) [23]. The PSD is described below as (1), where $x$ is the windowed raw EEG segment, $F(x)$ is the amplitude values after Fourier transform from the EEG segments and $fmin$ and $fmax$ are the minimum and maximum frequencies of given in Table III.

$$PSD = \sum_{i=fmin}^{fmax} F(x)_i \qquad (1)$$

A 5-second sliding window can not always cover the whole period of the low-frequency slow eye movements [21], so they can be difficult to detect it using STFT even with zeropadding. To address this issue, the PSD of slow eye movement was captured by applying Fourier transform over the entire EEG epoch.

*2) Statistics of spectral power:* Additional information can be extracted from the spectral power. Some features such as alpha rhythm and sleep spindle usually appear regularly. The duration of each sub-band was estimated by averaging PSDs over an EEG epoch. Larger averaged PSD is equivalent to better continuation.

Low averaged and median PSD values with a high maximum PSD value appears as an occasional feature, such as K complexes and vertex shape waves. Moreover, the standard derivation of each sub-band evaluated the frequency fluctuation.

*3) Time domain:* The EEG amplitude is usually lower than $100uV$, while the EMG, EOG and movement artifact amplitudes are often higher. The maximum and minimum amplitudes of the raw EEG signal reflect artifact information. As the Shannon entropy of raw EEG signal [24] is sensitive to the amplitude distribution, it can additionally be used to derive related features from signal synchronization or amplitude.

### C. Mixed neural network

The temporal physiological features as introduced in the previous section can serve as the input data which can feed into different classes of classification algorithms, e.g., logistic regression, supporting vector machine, etc. Unfortunately, the classification performance of these has been poor because they do not address the temporal pattern recognition challenge [25], [26], [27]. Some of the physiological features are strongly correlated, although they may be incomplete relative to currently accepted sleep physiological descriptions. Further exploration in the feature space is needed. As emphasized earlier, any given sleep stage depends not only on the features at the moment, but also on those that are highly correlated in the past. Such temporal dependency needs to be considered.

Motivated by these considerations, two classes of deep neural networks are introduced: multi-layer perception (MLP) and a recurrent neural network (RNN) to address completeness and temporal correlations, respectively. As illustrated in Fig. 2, after independent applications, MLP and RNN are concatenated in our model.

In the end, a softmax function is introduced for classification. In summary, the key idea is to use a mixture of neural networks to "learn" new features for classification. This type of mixture is termed as mixed neural network (MNN).

In Fig. 2, MNN specifies a modular structure for MLP, RNN and softmax respectively. For the MLP module, there are many latent parameters needed to be selected and tuned, such as the number of layers, number of units in each layer and the selection of nonlinear activation functions. A similar problem arises with the RNN module, as well; the structure of RNN needs to be specified a prior.

In our study, we choose a rectifier neural network as the candidate for MLP module and long-short term memory architecture for RNN module. These selections are mainly based on an *ad-hoc* tests across various combinations of these parameters, especially the number of layers and number of units across layers. Nevertheless, we provide an analysis on the selection of rectifier function as the candidate nonlinear activation function in MLP module.

*1) Rectifier Neural Network:* A rectifier neural network was used in the MLP module, where the function is described as follows:

$$f(x) = max(0, x). \qquad (2)$$

The rectifier network has proved to be able to optimize performance without any unsupervised pre-training on unlabelled data. It is well known that rectifying neurons performed better when the data is sparse compared to sigmoid and hyperbolic tangent neurons [18]. In our case, the EEG spectrum is a typical type of sparse data. First of all, only a few frequency bands will exist in any particular sleep stage. Secondly, the frequency activities are discontinuous, e.g., the alpha activity may only appear for about 50% of the epoch in stage W. Thirdly, different people may show different frequency amplitudes in same sleep stage.

In addition, the output is a linear function of the inputs, so the gradients from next RNN module are able to back-propagate well to all layers. This alleviate the "vanishing gradient" problem. As the spectrogram is used as input, the output of a rectifier neuron in the first hidden layer is the combination of PSDs features. Combinations of absent frequency bands output as zero with a rectifier, so their values will not effect the inputs of next layer. For discontinuous features and variation problem, the rectifier can represent whether these features exist by active or not.

Dropout is a popular technique for addressing overfitting in deep neural networks [11], [28]. The dropout method is able to train a large number of different networks while allowing all of the networks to share the same weights for the hidden neuron. It can be considered as another form of ensemble learning [11], and has similarities to autoencoders that we have described previously [4].

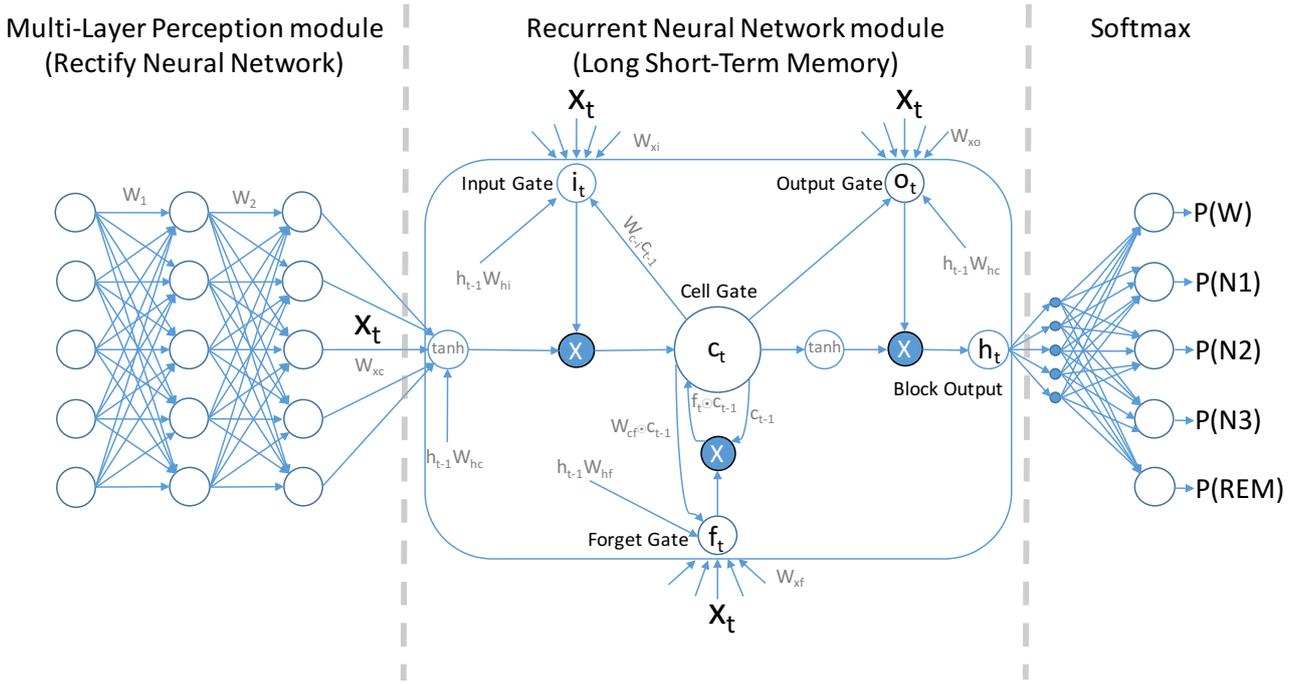

Fig. 2. Structure of mixed neural network. The spectrum features were input from the left, then we take advantage of rectifier neural network for detecting hierarchical feature and long short-term memory (LSTM) network for sequential data learning to optimize classification performance with single-channel recordings.

In our work here, we set the dropout probability from input layer to first rectifier layer, from first rectifier layer to second rectifier layer, and from second rectifier layer to RNN module as 20%, 50% and 50%, respectively. We found empirically that this combination of dropout probabilities achieved the best performance, which is similar to the dropout experimental study on MNIST dataset [11].

Without the rectifier neural network, the accuracy dropped by 3%. Moreover, in our experiment, we found the rectify activation function work better than sigmoid and hyperbolic tangent functions. It is also possible that this is caused by the non-linearities, e.g., when using sigmoid or hyperbolic tangent functions, even the input values are very high, the output will not change too much, because the output is close to 1.

*2) Long short-term memory:* Long short-term memory (LSTM) architecture is selected as the candidate in the RNN module. LSTM is well known to be capable of learning long-term dependencies problem [29], [19]. The advantage of LSTM is that it not only applied the current information to perform the present task, but also explicitly takes account to long-term information in the past which is a limitation of classic RNN architecture (the long-term dependency problem). Overall, the long term information memory is a key property of the LSTM architecture.

An illustrative example of long-term dependency problem in sleep stages classification arises when a long period of N2 sleep shifts to stage N1 for several EEG epochs. If sleep spindles appear, the probability of reverting to stage N2 is higher than the probability of progressing to stage N3.

In this study, LSTM is implemented by the following formulas, which is the vanilla architecture. It should be noted that the vanilla LSTM outperformed than any other variations as shown in [30]. The function $\sigma_h$ and $\sigma_c$ are the hyperbolic tangent activation function applied to the block output and cell gate, the other $\sigma$ functions are logistic sigmoid function. The element-wise multiplication (Hadamard product) of two vectors is denoted by $\odot$.

$$i_t = \sigma_i(x_t W_{xi} + h_{t-1} W_{hi} + w_{ci} \odot c_{t-1} + b_i) \quad \text{input gate}$$
$$f_t = \sigma_f(x_t W_{xf} + h_{t-1} W_{hf} + w_{cf} \odot c_{t-1} + b_f) \quad \text{forget gate}$$
$$c_t = f_t \odot c_{t-1} + i_t \odot \sigma_c(x_t W_{xc} + h_{t-1} W_{hc} + b_c) \quad \text{cell state}$$
$$o_t = \sigma_o(x_t W_{xo} + h_{t-1} W_{ho} + w_{co} \odot c_t + b_o) \quad \text{output gate}$$
$$h_t = o_t \odot \sigma_h(c_t) \quad \text{block output}$$

*3) Output module:* After LSTM, multinomial classification (softmax regression) which is widely used in various probabilistic multi-class classification problems, was applied as the output layer. The softmax regression is the generalization of logistic regression to multiple categories, which is to predict the probability of inputs $(x)$ belonging to each class $(y)$. The number of outputs of the softmax layer is equal to the number of classes; as there are 5 stages in our classification, the number of outputs is 5.

### D. Training the network

The MNN was trained by using stochastic gradient descent (SGD) [31] with a batch size of 500 examples and learning rate of 0.01 with momentum of 0.9. According to the softmax output and dropout of rectifier neural network, cross-entropy was used as the loss function without any kinds of weight decay.

## III. Experiment and Result

### A. Experimental Data

An open access dataset [32] was used to evaluate the proposed method. It includes data from 62 healthy subjects, aged from 23 to 73 years (29 men, 33 women). All recordings are from different subjects. Each EEG recording had an epoch duration of 30 seconds and recordings were scored by a single sleep expert following the AASM standard. Sleep stages are labeled as stage W, N1, N2, N3, REM or unknown. The unknown stage only exists at the start and the end of recordings, when the subjects are preparing to sleep or when the subjects are completing the recording. To evaluate the proposed method, only the "in-bed" part of the recording was taken, so the unknown stage was ignored.

The recording contains 20 EEG, 2 EOG, 1 ECG and 3 EMG channels, and all EEG channels are referential. This means that we are testing whether our method, which uses limited physiological data, approaches performance similar to that of the expert reader with a full set of bio-signal data. We used the derivation between F4 and EOG Left Horizon. They are placed near the hairline and outer-down canthus of left eye, they are placed on the skin with hair. The motivation for which being demonstration of a proof of principle for a convenient and comfortable to home-based EEG recording electrode configuration.

### B. Experimental Design

To evaluate the Mixed Neural Network, we tried several models for the rectifier neural network by varying the number of hidden layers (2 to 5), the number of hidden units in rectifying layer (200 to 800) and the number of hidden units of LSTM (200 to 1000). We also compared it with three representative classifiers, including the Support Vector Machine (SVM), Random Forest (RF) and Multilayer Perceptron (MLP). All classifiers used the same features best interpret comparisons across methods.

The SVM used radial basis function (RBF) kernel, with the kernel coefficient $gamma$ equals to 0.025. We set $C$ equal to 0.5 to regularize the estimation in order to avoid noisy features. Shrinking heuristic was also applied. The RF used 100 estimators, and the performance of a split was estimated as the mean square error. The number of features to consider when looking for the best split was equal to the square root of number of features. The nodes of RF where expanded until the "leaves" were pure or until the "leaves" contained less than 2 samples. Both SVM and RF avoid class imbalances by setting class weight. The MLP in this comparison used 2 layers rectifying neuron followed by softmax output layer, and the random dropout was applied during supervised training. The same approach was used with the feature processing layer of the Mixed Neural Network. All hyper-parameters were fine-tuned to achieve best performance. Oversampling was used to avoid the class imbalance problem when training the MLP.

We defined a rule for the SVM, RF and MLP. If the sequence length (SL) is 1, we used only the features of current EEG epoch to train the classifier. If sequence length is 2, we trained the classifier by using the features from current EEG epoch and previous 1 EEG epoch, and so on. In LSTM, sequence length is the number of examples considered for each output.

The classification performance was evaluated using widely-used indexes: macro F1-score (MF1) and overall accuracy (ACC). We generated the confusion matrix using the predictions from all of the cross-validation folds and we used this confusion matrix to calculate the overall accuracy and the macro F1-scores. The accuracy was computed from the sum of the diagonal elements divided by the total number of samples. The macro F1-score (MF1) was the mean of the per-class F1-scores of each sleep stage. In addition, the performances of each individual stage were evaluated by using the recall (RE), precision (PR) and F1-score (F1).

In this experiment, the training and testing data were from different subjects in order to limit overfitting and data dependence. K-fold cross-validation was adopted, $K$ was set to 31 for the 62 recordings, which means 2 recordings were used as a validation set and the other 60 recordings were used as a training set for each validation.

The reason of using a cross-validation is that, if we split our dataset into the training, validation and testing sets (e.g., 60%-20%-20%), we would have only 37, 12 and 13 patients for training, validation and testing respectively. Such small sets of subjects might not generalize to a larger population. We therefore used a 31-fold cross validation technique, in which we split the dataset into 31 subsets (each consists of 2 subjects). 30 subsets were used for training and the withheld subset was used for testing. This process was repeated 31 times, such that each of the 31 subsets was evaluated once.

To speed up the evaluation, Graphical Processing Units (GPU) acceleration was used to training the network. Training the networks for sequence length equal to 5 over cross validation takes about 2 days by using NVIDIA 630 GPU on a single machine. The code was implemented by Theano [33], [34].

### C. Result and Discussion

TABLE IV
Comparisons of different algorithms using F4-EOG (Left). SVM represents Support Vector Machine, RF represents Random Forest, MLP represents Multi-layer Perceptron, MNN represents Mixed Neural Network and SL represents Sequent Length

| SL | 1 | 2 | 3 | 4 | 5 | 6 |
|---|---|---|---|---|---|---|
| **SVM** | | | | | | |
| MF1 | 73.43 | **75.01** | 74.71 | 74.34 | 73.92 | 73.50 |
| ACC | 78.01 | **79.70** | 79.53 | 79.20 | 78.87 | 78.53 |
| **RF** | | | | | | |
| MF1 | 69.95 | 72.07 | **72.44** | 71.70 | 71.32 | 70.92 |
| ACC | 80.68 | 81.53 | **81.67** | 81.30 | 81.18 | 80.94 |
| **MLP** | | | | | | |
| MF1 | 74.11 | 76.71 | 76.80 | **77.23** | 76.71 | 76.23 |
| ACC | 78.17 | 80.47 | 80.81 | **81.43** | 81.41 | 81.37 |
| **MNN** | | | | | | |
| MF1 | 73.71 | 78.49 | 79.76 | 80.35 | **80.50** | 80.44 |
| ACC | 82.67 | 84.60 | 85.28 | 85.67 | **85.92** | 85.82 |

Table IV shows the macro F1-score (MF1) and overall accuracy (ACC) of different classifiers from cross-validation under different sequence length, while the boldface numbers indicated the best performance of a classifier.

We found that the performance of control group (SVM, RF and MLP) shows slight improvement when using the features from previous EEG epoch(s) compared with using only the features from current EEG epoch (sequence length=1). However, no continual improvement found; results even became worse as the sequence length increased. For example, the performance of SVM became worse when the sequence length increased from 2 to 5. We believe that this occurs because control group transfers the classification problem as a complex formula and not as a sequential model. In this experiment, SVM reached its best performance when using the features from the current and previous EEG epoch. RF and MLP gave their best performance by using features from 3 and 4 nearest EEG epochs, respectively. Moreover, RF has better accuracy compare with SVM and MLP, but its macro F1-score is the worst.

By contrast, both the macro F1-score and overall accuracy of the Mixed Neural Network showed continual improvements as the sequence length increased from 1 to 5. The result demonstrated that our network has ability to remember the stage information epoch-by-epoch. Even setting the sequence length to 1, its accuracy is still better than the control group. When its sequence length is set to 5, both overall accuracy and macro F1-score become significantly higher than other classifiers. However, we found that the performance decreased when the sequence length was greater than 5. This may be due to the fact that the sleep experts would not consider the PSGs more than 5 epochs when labelling the data according to AASM manual. Thus, an overlong sequence length may not improve the performance but may introduce noise during training. Table V shows its confusion matrix from cross-validation, the left column is the actual class labeled by sleep expert, and the top row is the predicted class calculated by the classifier.

In addition, the standard deviations of the accuracies in each fold of MNN are 0.0604, 0.06648, 0.06020, 0.05865, 0.0284, 0.0343 with a sequence length from 1 to 6. We found that the standard deviation became stable when the sequence length was greater than 5.

For more details, Table VI compared MNN with the control group across different sleep stages. This table lists the best performance of different classifiers.

The boldface numbers indicated the best situation across classifiers in control group, and then the bottom line list the improvement of MNN compare with the control group. It is clear not only that MNN has better overall accuracy and macro F1-score, but also that all F1-scores for individual sleep stages are better than those for the control group overall.

In our experiment, we tried to add fully connected layers between LSTM and softmax, and vary their hidden sizes, but no improvement was found. Specifically, more fully connected layers after LSTM gave poorer performance than the one without any extra layers. We also tried to use dropout to avoid the overfitting, but the performance was similar to the one without any extra layers. Actually, many LSTM-based applications, such as image captioning [35] and visual question answering [36] outputting the probabilities of every word in a vocabulary (i.e., very large number of output dimensions), also fed the activations from LSTM output to a softmax layer directly. We therefore decided not to add any extra layers before the LSTM.

We did not found better performance if we used more fully connected layers in the MLP module before LSTM. The reason is we already extracted the features from EEG signals. Adding unnecessary layers will complicate the network and lead to the overfitting problem. For instance, when we used 3 layers of MLP with the same dropout probabilities, the accuracy slightly decreased. When we set the dropout probabilities to a higher value, our model outputted a similar accuracy with our current architecture. We tried different numbers of hidden units from (200 to 800), and different activation functions such as rectifier, sigmoid, hyperbolic tangent and ramp. We found that the 2 layers of MLP with 300 hidden units and the rectifier activation function gave us the best sleep stage scoring performance.

We also explored different types of RNNs, such as vanilla RNN, long short-term memory (LSTM) and gated recurrent unit (GRU). We found that vanilla RNN gave a poorer performance compared to LSTM and GRU. Both LSTM and GRU were better in distinguishing sleep stages N1 and N2. We found that they gave similar performance, which is the same phenomenon observed in [30].

To explore whether MLP can learn a combination of features, we tried a model without MLP, and we found that the performance in distinguishing between "N2 and N3", and "W and REM" decreased (e.g., the overall accuracy decreased to 85.10%). This implied that the MLP was able to learn useful combination of the preprocessed features. For instance, the MLP might learn a common representation of sleep spindles appearing in sleep stage N2 and N3, and rapid eye movement in stage W and REM.

Compared with existing studies, Table VII shows their macro F1-scores, overall accuracy and the F1-scores of different sleep stages. It shows that our macro F1-score, and F1-score of stage N1 and N2 are significantly higher than the existing studies. That is because LSTM performs better when dealing with the continuation of N1 and N2. However, the F1-score of stage W is lower than most of existing studies. It is caused by poor detection of alpha rhythm from frontal lobe,

TABLE V
CONFUSION MATRIX FROM CROSS-VALIDATION USING MNN AND F4-EOG (LEFT) WHEN SEQUENCE LENGTH IS 5. THE LEFT COLUMN IS THE ACTUAL CLASS LABELED BY SLEEP EXPERT, AND THE TOP ROW IS THE PREDICTED CLASS CALCULATED BY THE CLASSIFIER.

ACC = 85.92% MF1 = 80.50%

| SL=5 | W | N1 | N2 | N3 | REM | RE | PR | F1 |
|------|------|------|-------|------|------|-------|-------|-------|
| W | 5022 | 577 | 188 | 19 | 395 | 80.95 | 88.49 | 84.55 |
| N1 | 407 | 2468 | 989 | 4 | 965 | 51.07 | 62.75 | 56.31 |
| N2 | 130 | 630 | 27254 | 1021 | 763 | 91.46 | 90.02 | 90.73 |
| N3 | 13 | 0 | 1236 | 6399 | 5 | 83.61 | 85.94 | 84.76 |
| REM | 103 | 258 | 609 | 0 | 9611 | 90.83 | 81.87 | 86.12 |

TABLE VI

COMPARISON BETWEEN OUR METHOD AND OTHER CLASSIFIERS ACROSS THE FIVE SCORING PERFORMANCE METRICS (PRECISION, RECALL, F1-SCORE, MACRO F1-SCORE, AND OVERALL ACCURACY) USING F4-EOG (LEFT)

| Method | SL | MF1 | ACC | W | | | N1 | | | N2 | | | N3 | | | REM | | |
|--------|----|-----|-----|-----|-----|-----|-----|-----|-----|-----|-----|-----|-----|-----|-----|-----|-----|-----|
| | | | | RE | PR | F1 | RE | PR | F1 | RE | PR | F1 | RE | PR | F1 | RE | PR | F1 |
| SVM | 2 | 75.01 | 79.70 | 84.14 | 73.79 | 78.63 | 59.76 | 41.14 | 48.73 | 78.78 | 94.81 | 86.06 | 91.52 | 75.03 | **82.46** | 80.20 | 78.20 | 79.19 |
| RF | 3 | 72.44 | 81.67 | 77.73 | 78.70 | 78.21 | 23.60 | 68.69 | 35.13 | 93.03 | 83.49 | **88.00** | 76.05 | 87.78 | 81.50 | 82.65 | 76.35 | 79.38 |
| MLP | 4 | **77.23** | **81.43** | 83.30 | 82.62 | **82.95** | 58.47 | 49.11 | **53.38** | 78.89 | 94.77 | 86.10 | 95.09 | 69.70 | 80.44 | 88.11 | 78.91 | **83.26** |
| MNN | 5 | 80.50 | 85.92 | 80.95 | 88.49 | 84.55 | 51.07 | 62.75 | 56.31 | 91.46 | 90.02 | 90.73 | 83.61 | 85.94 | 84.76 | 90.83 | 81.87 | 86.12 |
| | | +3.27 | +4.49 | | | +1.59 | | | +2.93 | | | +2.73 | | | +2.30 | | | +2.86 |

TABLE VII

COMPARISON BETWEEN OUR METHOD AND THE LITERATURE ACROSS THE THREE SCORING PERFORMANCE METRICS (F1-SCORE, MACRO-F1 SCORE, AND OVERALL ACCURACY)

| Study | Channel | MF1 | ACC | F1 | | | | |
|-------|---------|-----|-----|-----|-----|-----|-----|-----|
| | | | | W | N1 | N2 | N3 | REM |
| [4] | Fpz-Cz | 73.73 | 78.94 | 71.58 | 47.04 | 84.60 | 84.03 | 81.40 |
| [5] | Cz-Pz | 74.74 | 82.57 | 91.45 | **47.62** | 82.59 | 74.21 | 77.81 |
| [6] | C3-A2/Pz-Oz | 68.11 | **83.60** | **93.62** | 15.29 | 78.25 | 71.45 | 81.96 |
| [7] | Cz-Pz | 72.36 | 82.94 | 85.95 | 20.86 | 84.78 | 84.28 | 85.95 |
| [14] | Fp1-Fp2 | **76.59** | 81.65 | 74.82 | 47.04 | **86.94** | **87.89** | **86.31** |
| MNN | F4-EOG Left | 80.50 | 85.92 | 84.55 | 56.31 | 90.73 | 84.76 | 86.12 |
| | | +3.91 | +2.32 | -6.90 | +8.69 | +3.79 | -3.13 | -0.19 |

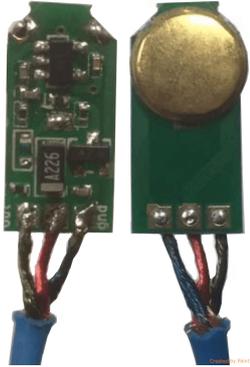

Fig. 3. Active ultra-high impedance electrode design (Left: circle side; Right: skin contact side)

but it outperformed two studies using EEG from central lobe or frontal lobe.

## IV. CONCLUSION AND DISCUSSION

In this paper, the proposed Mixed Neural Network and the corresponding training method work well for sleep stages classification problem compared with SVM, RF and MLP. Moreover, the proposed method only uses EEG signal from a single pair of electrodes positioned comfortably over hairless skin.

However, in terms of convenience, wearing the F4 channel near the hair line is imperfect. Other frontal EEG channels such as Fp2 and Fpz are easier to wear, but these channels have lesser information about stage W, N1, N2 and N3 compared with the F4 channel due to the longer distance to central lobe,

TABLE VIII

CONFUSION MATRIX FROM CROSS-VALIDATION USING MNN AND *Fp2-EOG (Left)* WHEN SEQUENCE LENGTH IS 5

| SL=5 | ACC = 83.35% MF1 = 76.97% | | | | | | | |
|------|-----|-----|-----|-----|-----|-----|-----|-----|
| | W | N1 | N2 | N3 | REM | RE | PR | F1 |
| W | 4604 | 795 | 294 | 32 | 479 | 74.21 | 86.04 | 79.69 |
| N1 | 405 | 2208 | 1292 | 9 | 919 | 45.69 | 57.08 | 50.75 |
| N2 | 208 | 605 | 27199 | 897 | 889 | 91.28 | 86.65 | 88.91 |
| N3 | 24 | 1 | 1689 | 5936 | 3 | 77.56 | 86.22 | 81.66 |
| REM | 110 | 259 | 914 | 11 | 9287 | 87.77 | 80.22 | 83.83 |

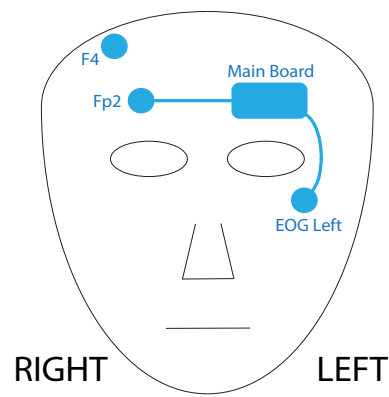

Fig. 4. Proposed home-care sleep monitoring configuration. Only the EEG signal from forehead is required.

as Table II describes. To evaluate these channels, Fp2-EOG left was evaluated by using the Mixed Neural Network and the same feature extraction algorithm with F4-EOG left. The result on Table VIII shows lower accuracy and macro F1-score compare with F4-EOG left. However, the result still better than other classifies in control group when they used EEG from F4-EOG left.

Fig. 4 illustrates the idea of home-care sleep monitoring system using Fp2-EOG left and dry electrode as Fig. 3 shows. The main board contains amplifiers, an analog to digital converter and wireless transmission module. The driven-right leg (DRL) can be placed on anywhere such as the back side of main board. With this structure, the device can be designed as a sleep mask, then movement during sleep would not lead to uncomforted feeling and noisy signal. This is a potential hardware approach for the proposed method in this paper.

## Appendix A
## Confusion matrices from cross-validation of different classifiers using same features

In order to make a fair comparison, these algorithms used same features as well as the proposed method on Table V, and same feature extraction algorithm as well as Table VIII. Moreover, as SVM, RF and MLP have their best performance when sequence lengths are 2, 3 and 4 respectively as Table IV shows, only the confusion matrices with best performance are shown.

### TABLE IX
### SVM USING F4-EOG (LEFT) WHEN SEQUENCE LENGTH IS 2

| | | | | | | ACC = 79.70% MF1 = 75.01% | | |
|---|---|---|---|---|---|---|---|---|
| SL=2 | W | N1 | N2 | N3 | REM | RE | PR | F1 |
| W | 5369 | 574 | 73 | 17 | 348 | 84.14 | 73.79 | 78.63 |
| N1 | 713 | 2891 | 430 | 8 | 796 | 59.76 | 41.14 | 48.73 |
| N2 | 536 | 2266 | 23479 | 2302 | 1219 | 78.78 | 94.81 | 86.06 |
| N3 | 132 | 2 | 512 | 7004 | 3 | 91.52 | 75.03 | 82.46 |
| REM | 526 | 1295 | 270 | 4 | 8486 | 80.20 | 78.20 | 79.19 |

### TABLE X
### RF USING F4-EOG (LEFT) WHEN SEQUENCE LENGTH IS 3

| | | | | | | ACC = 81.67% MF1 = 72.44% | | |
|---|---|---|---|---|---|---|---|---|
| SL=3 | W | N1 | N2 | N3 | REM | RE | PR | F1 |
| W | 4914 | 202 | 606 | 19 | 581 | 77.73 | 78.70 | 78.21 |
| N1 | 743 | 1141 | 1709 | 1 | 1241 | 23.60 | 68.69 | 35.13 |
| N2 | 250 | 160 | 27724 | 790 | 878 | 93.03 | 83.49 | 88.00 |
| N3 | 17 | 0 | 1808 | 5820 | 8 | 76.05 | 87.78 | 81.50 |
| REM | 320 | 158 | 1358 | 0 | 8745 | 82.65 | 76.35 | 79.38 |

### TABLE XI
### MLP USING F4-EOG (LEFT) WHEN SEQUENCE LENGTH IS 4

| | | | | | | ACC = 81.43% MF1 = 77.23% | | |
|---|---|---|---|---|---|---|---|---|
| SL=4 | W | N1 | N2 | N3 | REM | RE | PR | F1 |
| W | 5218 | 550 | 88 | 30 | 378 | 83.30 | 82.62 | 82.95 |
| N1 | 609 | 2826 | 533 | 12 | 853 | 58.47 | 49.11 | 53.38 |
| N2 | 258 | 1658 | 23508 | 3117 | 1258 | 78.89 | 94.77 | 86.10 |
| N3 | 12 | 4 | 357 | 7277 | 3 | 95.09 | 69.70 | 80.44 |
| REM | 218 | 717 | 319 | 4 | 9323 | 88.11 | 78.91 | 83.26 |

We need to point out these existing single-channel based studies used different dataset for evaluation, so it is not suitable to compare them directly by using accuracy. However, the recall, precision and F1-score can illustrate the reliability of algorithms, especially for stage N1. The confusion matrices are borrowed from their papers.

### TABLE XII
### COMPLEX MORLET WAVELETS FROM FPZ-CZ USING STACKED SPARSE AUTOENCODERS [4]

| | | | | | | ACC = 78.94% MFI = 73.73% | | |
|---|---|---|---|---|---|---|---|---|
| | W | N1 | N2 | N3 | REM | RE | PR | F1 |
| W | 2744 | 441 | 34 | 23 | 138 | 81.18 | 64.01 | 71.58 |
| N1 | 471 | 1654 | 262 | 6 | 366 | 59.91 | 38.73 | 47.04 |
| N2 | 621 | 1270 | 13696 | 1231 | 760 | 77.92 | 92.53 | 84.60 |
| N3 | 143 | 7 | 469 | 4966 | 6 | 88.82 | 79.74 | 84.03 |
| REM | 308 | 899 | 340 | 0 | 6164 | 79.94 | 82.92 | 81.40 |

### TABLE XIII
### CWT AND RENYI'S ENTROPY FROM CZ-PZ USING RANDOM FOREST CLASSIFIER [5]

| | | | | | | ACC = 82.57% MFI = 74.74% | | |
|---|---|---|---|---|---|---|---|---|
| | W | N1 | N2 | N3 | REM | RE | PR | F1 |
| W | 2407 | 89 | 111 | 38 | 40 | 89.65 | 93.33 | 91.45 |
| N1 | 56 | 185 | 52 | 8 | 48 | 53.01 | 43.22 | 47.62 |
| N2 | 69 | 85 | 1897 | 174 | 131 | 80.52 | 84.76 | 82.59 |
| N3 | 14 | 9 | 86 | 482 | 3 | 81.14 | 68.37 | 74.21 |
| REM | 33 | 60 | 92 | 3 | 719 | 79.27 | 76.41 | 77.81 |

### TABLE XIV
### MULTISCALE ENTROPY AND AUTOREGRESSIVE MODELS FROM C3-A2 OR PZ-OZ USING LINEAR DISCRIMINANT ANALYSIS [6]

| | | | | | | ACC = 83.60% MFI = 68.11% | | |
|---|---|---|---|---|---|---|---|---|
| | W | N1 | N2 | N3 | REM | RE | PR | F1 |
| W | 1849 | 87 | 59 | 4 | 11 | 91.99 | 95.31 | 93.62 |
| N1 | 69 | 24 | 12 | 3 | 20 | 18.75 | 12.90 | 15.29 |
| N2 | 15 | 45 | 669 | 165 | 59 | 70.20 | 88.38 | 78.25 |
| N3 | 0 | 1 | 1 | 224 | 0 | 99.12 | 55.86 | 71.45 |
| REM | 7 | 29 | 16 | 5 | 334 | 85.42 | 78.77 | 81.96 |

### TABLE XV
### SPECTRAL / TEMPORAL FEATURE EXTRACTION FROM CZ-PZ USING FUZZY CLASSIFICATION [7]

| | | | | | | ACC = 82.94% MFI = 72.36% | | |
|---|---|---|---|---|---|---|---|---|
| | W | N1 | N2 | N3 | REM | RE | PR | F1 |
| W | 1609 | 136 | 134 | 20 | 52 | 82.47 | 89.74 | 85.95 |
| N1 | 88 | 85 | 41 | 1 | 24 | 35.56 | 14.76 | 20.86 |
| N2 | 37 | 250 | 4534 | 467 | 139 | 83.55 | 86.05 | 84.78 |
| N3 | 0 | 0 | 369 | 2303 | 0 | 86.19 | 82.46 | 84.28 |
| REM | 59 | 105 | 191 | 2 | 1749 | 83.05 | 89.05 | 85.95 |

TABLE XVI
Fp1-Fp2 USING HIERARCHICAL DECISION TREE [14] ON WELL-RESTED SUBJECTS.

ACC = 81.65% MF1 = 76.59%

|     | W    | N1  | N2   | N3   | REM  | RE    | PR    | F1    |
| --- | ---- | --- | ---- | ---- | ---- | ----- | ----- | ----- |
| W   | 1046 | 147 | 26   | 3    | 102  | 79.00 | 71.06 | 74.82 |
| N1  | 252  | 795 | 172  | 7    | 305  | 51.93 | 42.99 | 47.04 |
| N2  | 35   | 599 | 5390 | 292  | 106  | 83.93 | 90.16 | 86.94 |
| N3  | 21   | 92  | 346  | 2806 | 0    | 85.94 | 89.94 | 87.89 |
| REM | 118  | 216 | 44   | 12   | 2846 | 87.95 | 84.72 | 86.31 |

## ACKNOWLEDGMENT

The authors would like to thank Center for Advanced Research in Sleep Medicine, University of Montreal for providing the data, especially like to thank Christian O'reilly for her helpful answers. PMM gratefully acknowledges support from the Edmond J. Safra Foundation and Lily Safra and the Imperial College Healthcare Trust Biomedical Research Centre. PMM is an NIHR Senior Investigator. The authors are grateful to Pan Wang for her valuable comments and suggestions for design of the sleep mask.

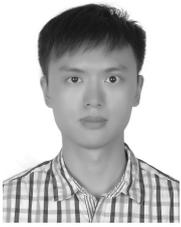

**Hao Dong** is currently a Ph.D. candidate at Data Science Institute and Department of Computing, Imperial College London. His research interests are in the areas of machine learning theory and biomedical engineering, especially with EEG and fMRI data.

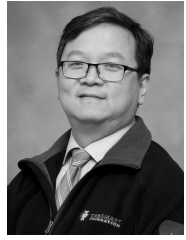

**Yike Guo** is currently a Professor of Computing Science in the Department of Computing at Imperial College London. He is the founding Director of the Data Science Institute at Imperial College, as well as leading the Discovery Science Group in the department.

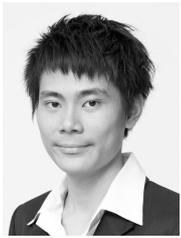

**Akara Supratak** is currently a Ph.D. candidate at Data Science Institute, Department of Computing, Imperial College London. His research interests includes biomedical engineering, software engineering and machine learning, especially with time series data.

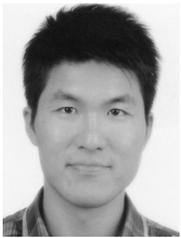

**Wei Pan** is now a Research Associate at Data Science Institute, Imperial College London. He is interested in machine learning theory and applications in data driven research especially in nonlinear time series modeling and prediction.

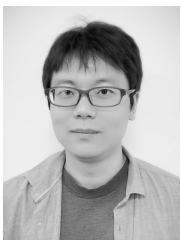

**Chao Wu** received his PhD degree from Zhejiang Univeristy, China. From 2011, he is now a Research Associate in the Discovery Science Group working on Elastic Sensor Network. His main research focus on data modelling and sensor network.

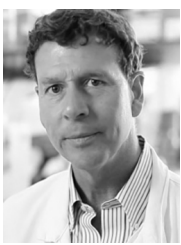

**Paul M. Matthews** OBE, MD, DPhil, FRCP, FMed-Sci is Head of the new Division of Brain Sciences in the Department of Medicine of Imperial College, London. His research is noted for innovative translational applications of clinical imaging for the neurosciences. This has developed with exploitation of the powerful synergies between the physical and quantitative sciences and medicine.